\documentclass[a4paper,10pt,oneside]{article}
\usepackage[latin1]{inputenc}
\usepackage{amsmath}
\usepackage{amsfonts}
\usepackage{amssymb}
\usepackage{graphicx}
\usepackage{footnote}
\usepackage{float}
\usepackage{subcaption}
\DeclareGraphicsExtensions{.bmp,.png,.pdf,.jpg,.eps}
\usepackage{physics}
\usepackage{mathrsfs}
\usepackage{cite}
\usepackage[colorlinks,urlcolor=blue,citecolor=blue]{hyperref}

\advance \textheight by \topskip
\usepackage{amsmath}
\usepackage[english]{babel}
\makeatletter
 \@addtoreset{equation}{section}
\makeatother
\usepackage[latin1]{inputenc}
\usepackage{amsmath}
\numberwithin{equation}{section}

\advance \textheight by \topskip
\usepackage{amsmath}
\usepackage[english]{babel}
\DeclareMathAlphabet\mathbfcal{OMS}{cmsy}{b}{n}
\DeclareMathAlphabet{\boldmathe}{T1}{cmr}{bx}{it}

\def\be{\begin{equation}}
\def\ee{\end{equation}}

\def\R{\mathbb R}

\def\C{\mathbb C}

\def\I{\mathbb{I}}

\def\be{\begin{equation}}
\def\ee{\end{equation}}

\def\R{\mathbb R}

\def\C{\mathbb C}

\def\I{\mathbb{I}}

\def\be{\begin{equation}}
\def\ee{\end{equation}}

\def\R{\mathbb R}

\def\C{\mathbb C}

\def\I{\mathbb{I}}

\def\Ti{\text{i}}
 \newcommand{\sfrac}[2]{{\textstyle\frac{#1}{#2}}}

\usepackage{todonotes}

\begin{document}

\begin{center}
{\LARGE \bf
Non-Hermitian superintegrable systems\\
}
\vspace{6mm}
{\Large Francisco Correa$^a$, Luis Inzunza$^{a}$ and Ian Marquette$^b$ 
}
\\[6mm]
\noindent ${}^a${\em 
Instituto de Ciencias F\'isicas y Matem\'aticas\\
Universidad Austral de Chile, Casilla 567, Valdivia, Chile}\\[3mm]
\noindent ${}^b${\em
School of Mathematics and Physics, \\
The University of Queensland
Brisbane,  QLD 4072, Australia}
\vspace{12mm}
\end{center}

\begin{abstract}
A non-Hermitian generalisation of the Marsden--Weinstein reduction method is introduced to construct families of quantum $\mathcal{PT}$-symmetric superintegrable models over an $n$-dimensional sphere $S^n$. The mechanism is illustrated with one- and two-dimensional examples, related to $u(2)$ and $u(3)$ Lie algebras respectively, providing new quantum models with real spectra and spontaneous $\mathcal{PT}$-symmetric breaking. In certain  limits, the models reduce to known non-Hermitian systems and complex extensions of previously studied real superintegrable systems. 	

\end{abstract}

\section{Introduction}

One of the most well-known methods for the construction of integrable Hamiltonian systems is the Marsden--Weinstein reduction procedure \cite{mw,Marsden}. Starting from a free Hamiltonian defined on a higher dimensional homogeneous space, a desired set of symmetries from a Lie group is used to reduce the dimension of the model allowing different classes of potential interactions to appear. The reduced phase space is the quotient between the original phase space (symplectic manifold) and a symplectic symmetry group of Abelian transformations. Using the so-called momentum map \cite{Marsden}, an automorphism between vector fields and phase space functions that preserves algebraic relations, the reduction generates integrals of motion based on the Lie symmetries and hence the integrability properties of the models. Well-known applications of Hamiltonian reductions are the heavy top and the two-body problem, where
the dimensional reduction is due to translational and rotational invariance \cite{mw,Marsden, babelon}. There are many mechanical systems where such techniques are useful, like rigid-elastic bodies \cite{mw2} and the Calogero and Neumann models \cite{Marsden, babelon, calop}. But what also makes the Marsden--Weinstein procedure powerful is that it can be used to find and to 
construct new superintegrable systems having separation of variables in the Hamilton--Jacobi and Laplace--Beltrami equations \cite{InSys1,InSys3,InSys15,InSys2}. Starting from a free particle in a homogeneous space invariant under pseudo-unitary transformations, the reduction takes place with the transformations generated by the so-called maximal Abelian subalgebras (MASAs) in the symmetry Lie algebra \cite{MASAs,MASAs2,masa1}. The reduced system has an effective potential whose structure results directly from the choice of the MASAs, and the second order integrals of motion of the model correspond to elements in the invariant sector of the enveloping algebra. Thus, the reduced systems with $n$ degrees of freedom are ``maximally" superintegrable, there are $2n-1$ functionally independent integrals of motion \cite{reviewsuper}. One of the standard examples of Hamiltonian models obtained by these reduction procedures is the generic model on the $2$-sphere,
\begin{equation}\label{example}
 H=p_1^2+p_2^2+p_3^2+\frac{k_1^2}{s_1^2} +\frac{k_2^2}{s_2^2} + \frac{k_3^2}{s_3^2} ,\qquad s_1^2+s_2^2+s_3^2=1 \,,
\end{equation}
where the generalisation to the $n$-sphere is straightforward. This system
 nicely illustrates the fact that superintegrable models possessing second order integrals of motion can be solved
  by separation of variables \cite{reviewsuper}. Together with many other systems constructed via the MASAs approach, these models have been extensively  studied in the literature, e.g. from the point of view of
   contractions \cite{contra, mil14}, intertwining operators \cite{cal06, cal09,cal09v2}, Wilson polynomials \cite{kmp}, quadratic algebras \cite{post11, vin19, kuru20} and also for an algebraic approach independent 
of the realisations \cite{CorOlMar},  to name just a few. 
Although the Marsden--Weinstein scheme has been studied
beyond mechanical systems, 
up to the authors knowledge, there is no a systematic approach using this construction for superintegrable non-Hermitian Hamiltonians with real spectra. The main idea of the present article is to fill this gap by designing new complex superintegrable solvable models. The appearance of complex potentials in the search for superintegrable and separable systems is, however, not new. The separability in the Hamilton-Jacobi and the Laplace-Beltrami equations has a long history of contributions that have been rigorously studied by several approaches \cite{reviewsuper, boyer, Miller, kalnins}. In this sense, Ref.  \cite{recent} provides a recent overview of the topic with a long list of models admitting separability, which includes the classification of orthogonal separable coordinate systems leading to some complex potentials. This idea was explored in a series of papers  \cite{comp1, comp2, comp3, comp4}, revealing complex superintegrable Hamiltonians depending on three independent parameters. In this scenario, the novelty of our approach is threefold. First, the idea is to find complex potentials under the group theoretical approach provided by the MASAs, rather than the separability in coordinate systems in the aforementioned works. Second, it is possible to generate new potentials depending in more than three independent parameters, generalising some of the cases studied in  \cite{comp2, comp3, comp4}. Finally, the present motivation takes into account the existence of ${\cal PT}$-symmetries and their peculiarities. Of course, before the appearance of non-Hermitian Hamiltonians in the scientific scene, the construction of superintegrable models using the MASAs \cite{InSys1,InSys3}  required the hermiticity condition in the potential. Nevertheless, 
it has been known for 
three decades that the reality of the quantum spectrum can be explained not only by the standard textbook approach but also, under certain conditions, from a non-Hermitian point of view \cite{ali, znojilconf, rev}. 
The most typical example is when a non-Hermitian Hamiltonian is symmetric under the simultaneous action of the parity and time reversal operators (${\cal PT}$)
 and the Hamiltonian wavefunctions are also eigenfunctions of ${\cal PT}$. 
 So far, these ideas are being applied 
 in a wide range of research areas, including optics, condensed matter physics, acoustics, field theories and quantum mechanics, and the list is still growing \cite{qscs,bookpt}. 
 Integrable systems are not the exception and also admit different types of non-Hermitian extensions or ${\cal PT}$-symmetric deformations. 
 For example, in classical integrable systems admitting soliton solutions,
  ${\cal PT}$-symmetry is used to deform well-known models \cite{ptsol1, ptsol2}, to construct new types of integrable models with novel solitons solutions in non-local form \cite{nonlocal, hirota} and to give a physical meaning to solitons with divergent charges, making them real \cite{regsol}. The solitons and non-linear equations with non-Hermitian properties became a whole field of research in their own right, still under development
    \cite{konotop2016nonlinear, andreasreview}. Both classical and quantum integrable many-particle systems  have also been 
    studied
     in terms of non-Hermitian extensions. One of the best representatives,
     the Calogero models were among the first integrable models to be extended to complex generalisations \cite{cal1, cal2, cal3} but
which have also been studied in a more sophisticated way via the complexification of root systems \cite{roots}, see \cite{reviewcal} for an updated list of references.  Quantum non-Hermitian Hamiltonians in terms of symmetry generators (or creation and annihilation operators) give rise to the Swanson model \cite{Swanson} and its extensions \cite{AsFr} including supersymmetric ones \cite{InzPly10}. Recently, complex magnetic fields have also attracted the attention from the point of view of superintegrability, see for instance \cite{complex1} and in  graphene systems with non-Hermitian Dirac-Weyl  Hamiltonians \cite{complex2}. So far there are several approaches to introduce non-Hermitian deformations, one way considers the complexification of coordinates, field or parameters, while another form modifies the underlying symmetry structure via the generators or the root systems. The idea presented here belongs more to the second one but is slightly different, the non-Hermitian extensions are introduced before the Marsden-Weinstein reductions, at the level of the MASAs, allowing a large class of complex parameters. \newline

The outline of the paper is as follows. Section \ref{SecTheo} reviews the Marsden--Weinstein approach with MASAs described in Refs. \cite{InSys1,InSys3,InSys15,InSys2}. The construction of non-Hermitian superintegrable models living in the $S^{n}$ sphere is then introduced together with the description of the second order integrals of motion in terms of the symmetry generators. Section \ref{SecS1} deals with a two-parameter family of complex models on a circle, where special cases lead to well-known $\mathcal{PT}$-symmetric quantum systems. Non-Hermitian examples on the 2-sphere are studied in Section \ref{SecS2}, providing a rich $\mathcal{PT}$-symmetry structure with an exact and broken phases depending on parameter values. Next, additional examples are given which turns out to be related with $\mathcal{PT}$-extensions of MASAs as non-compact Cartan, orthogonally decomposable, and 
the nilpotent subalgebras. They also generalize previous results of complex separable systems in the 2-sphere.  Finally, discussion and outlook are provided in Section \ref{SecCon}.

\section{Complex systems in the sphere}
\label{SecTheo}
The idea is based on the Marsden--Weinstein dimensional
reduction scheme \cite{mw,Marsden} applied to the geodesic motion of a free particle in a homogeneous space  \cite{InSys1,InSys3,InSys15,InSys2}.
By means of a suitable canonical transformation,  some of the coordinates do not appear explicitly in the resulting Hamiltonian. Those coordinates are called ignorable and the corresponding canonical momenta are integrals of motion. In this way, the entire phase space is reduced to a region where the ignorable variables vanish and their momenta can be treated as coupling constants in an emerging potential term. The main ingredients in such canonical transformations are the MASAs
 \cite{MASAs,MASAs2,masa1} 
 that generate a subgroup of isometric transformations and their rank fixes the number of the ignorable variables in the system \cite{MASAs2}. The reduction method for constructing real superintegrable Hamiltonian systems takes a free particle living in a $n$-dimensional sphere, represented in terms of the complex coordinates 
\begin{equation}
\label{SphericalRec1Real}
y^\mu \overline{y}_\mu =1\,,
\qquad y_\mu \in \C\,,\qquad
\mu=1,\ldots,n\,,
\end{equation}
where $y$ denotes
a $n$-dimensional column vector. In this parametrisation, the Hamiltonian of the system reads
\begin{equation}
\label{Ham1}
H=\frac{2}{m}{P}_\mu \overline{P}^\mu \, ,
\end{equation} 
obtained from the geodesic Lagrangian $L=\frac{m}{2}g_{\mu\nu}\dot{y}\dot{\overline{y}}$ with Poisson brackets $\{y^\mu ,P_\nu\}_{\text{PB}}=\delta_{\nu}^\mu$ and $P_\mu y^\mu +\overline{y}_\mu \overline{P}^\mu =0\,$ as  direct consequence of the geometrical constraint (\ref{SphericalRec1Real}).  As the isometries of the homogeneous space (\ref{SphericalRec1Real}) are generated by the unitary group $U(n)$, the coordinate transformations $y'=By$ are given by $B$ as an unitary matrix. The symmetry group contains an Abelian subgroup of transformations, generated by the MASAs
and  their generators, denoted by $Y_\mu$, can be realized in terms of symmetric and purely imaginary matrices. Thus,  the reduction scheme establishes a canonical transformation of the form \cite{InSys1},
 \begin{eqnarray}  \label{trans1}
&y^\mu=B(x)^\mu_\nu s^\nu \, , \qquad \overline{y}^\mu=\overline{B}(x)^\mu_\nu s^\nu\,, & \\
 \label{trans2}
 & B(x)=\exp(x^{\mu} Y_\mu )\,,\quad
 [{Y}_\mu,{Y}_\nu]=0 \, , \qquad Y_\mu=-\overline{Y}_\mu=Y_\mu^T\,,
\end{eqnarray}
where $x^\mu$ and $s^\mu$ are real coordinates. The canonical transformation converts the two new sets of $n$-dimensional canonical momenta vectors $p_s$ and $p_x$, associated with the conjugate variables $s^\mu$ and $x^\mu$ respectively,  
\begin{equation}\notag
\left(\begin{array}{c}
\overline{P} \\
P
\end{array}\right)=(J^{-1})^{T}
\left(\begin{array}{c}
p_{x}\\
p_{s}
\end{array}\right)\,,
\end{equation}
where $J$ is the Jacobian matrix. The conditions in the $Y_\mu$ generators immediately imply 
 $\overline{B}=B^{-1}$ defining the $s^\mu$ variables over the $n$-dimensional sphere
\begin{equation}\label{sphere}
 s^\mu s_\mu=1\,,\qquad p_s^T s=0\, .
\end{equation}
The transformations (\ref{trans1}) and (\ref{trans2})  lead to the Hamiltonian (from now on, the mass is set to $m=1/2$),
\begin{equation}
\label{genrealcase}
H=p_s^T p_s+V(s,p_x)\,, \qquad V(p_x,s)= p_x^T {\cal V}^{-1} p_x\,,
\end{equation}
where ${\cal V}$ is, by construction, a matrix with real entries, defined in terms of the $s^\mu$ coordinates and MASAs generators,
\begin{equation}
\label{ComW}
\qquad
{\cal V}_{\mu\nu}=-({Y}_{\mu})_{\alpha}^\rho({Y}_{\nu})_{\rho}^\beta s^\alpha s_\beta\, ,  \qquad \Im  \mathcal{V}=0 \, .
\end{equation}
The above procedure
eliminates the presence of the ignorable variables $x^\mu$ in the potential and the phase space is reduced with the identification of the hyper-surface $x^\mu=0$ and $p_x=k=(k_1,k_2,\cdots, k_n) \in \R^n$ as new coupling constants. All together with (\ref{sphere}) defined the Hamiltonian (\ref{genrealcase}) living in the $S^{n}$ sphere with a non-trivial potential $V(p_x,s)=V(k,s)$.  More details about this construction are given in the Appendix \ref{AppA}, see also \cite{InSys1}. As we shall see below, this method generates integrals of motion in such a way the reduced system is superintegrable. Indeed, several extensions have been studied for different type of isometry groups \cite{InSys3,InSys15,InSys2}, as well as their relation with intertwining operators and supersymmetry \cite{cal06,cal09} and more recently from a pure algebraic approach \cite{CorOlMar}. However, the natural restriction of a real valued potential  (\ref{ComW}), imposes strong constraints on the different choices of the MASAs,  taking  into account that for larger symmetry groups more classes, and therefore possibilities, do appear. For instance, for rank three algebras, several of the integrable models obtained include terms 
with negative kinetic signature \cite{InSys2,InSys3}, making more difficult to connect those systems with physical backgrounds. The following construction consists in the generalisation of the method known so far, that overcomes those issues but also allows to find new families of superintegrable systems by making the potential non-Hermitian but ${\cal PT}$-symmetric. The generalisation start with replacing the homogeneous $2n$-dimensional space (\ref{SphericalRec1Real}) into new independent complex coordinates $q^\mu$ and $r^\mu$
\begin{align}\label{ext}
(y,\bar{y}) \rightarrow (q,r), \qquad (P, \overline{P}) \rightarrow (p_q, p_r) \, .
\end{align}
Naturally, the resulting configurational space admits a larger class of isometries and coordinate transformations in terms of a symmetric complex matrix, $B$ 
 \begin{eqnarray}
&\label{ComplexSphere}
q^T r=
q^\mu r_{\mu}=1\,, \quad q'=B q \, \quad r'=B^{-1}r \, .
\end{eqnarray}
In comparison with the original Hamiltonian (\ref{Ham1}), it is easy to see that the new one
can be treated as an analytic continuation of a free $2n$-dimensional one 
\begin{eqnarray}\label{newone}
H=4p^T_{q}p_r=4(p_+^T p_++p_-^T p_-), \quad p_\pm=p_q\pm \Ti p_r \, ,
\end{eqnarray}
although their interpretation does not require a deeper justification beyond that establishes an extension to construct new superintegrable models. From the new initial Hamiltonian (\ref{newone}), the reduction scheme (\ref{trans1}) and (\ref{trans2}) is extended into
 \begin{eqnarray}
&q^\mu=B(x)^\mu_\nu s^\nu \, , \qquad r^\mu={B^{-1}}(x)^\mu_\nu s^\nu & \\
 \label{transa}
 & B(x)=\exp(x^{\mu} Z_\mu )\,,\quad
 [{Z}_\mu,{Z}_\nu]=0 \, ,\qquad Z_\mu=Z_\mu^T\ , \qquad Z_\mu \in \C \, ,
\end{eqnarray}
where in this case the complex Abelian matrices $Z_\mu$ can be complex linear combination of all possible MASAs generating the $u(n)$ algebra $Z_\mu=\sum a_\mu Y_\mu$ . This new choice, not allowed in the previous approaches \cite{InSys1,InSys2}, implies that the Hamiltonian (\ref{newone}) and the dynamical restriction take 
a similar form as (\ref{genrealcase}), but the potential $V$ is changed due to the complex valued 
matrix,
\begin{eqnarray}
\label{ComV}
{\cal V}_{\mu\nu}=-({Z}_{\mu})_{\alpha}^\rho({Z}_{\nu})_{\rho}^\beta s^\alpha s_\beta\,.
\end{eqnarray}
The ignorable coordinates remain the same as before $x_\rho$, as well as, the reduced phase space. Obviously this extension contains the real scheme as a special case, but also admits new configurations in terms of $\mathcal{PT}$-symmetric superintegrable systems from (\ref{ComV}). The choice of the $\mathcal{PT}$-symmetric operator is not always unique \cite{ali} and Hamiltonians could have simultaneously different anti-linear commuting operators. In this setup, it is enough to demand the invariance of the potential $\mathcal{PT}(\mathcal{V})=\mathcal{V}$ using an operator like
\begin{equation}\label{ptgena}
\mathcal{T}: \Ti\rightarrow -\Ti\,,\qquad
\mathcal{P}: s^{\mu}\rightarrow \mathcal{P}^{\mu}_\nu s^\nu, \qquad  \mathcal{P}= \mathcal{P}^{-1}=\mathcal{P}^{T} \, ,
\end{equation}
which also preserves the kinetic term. The invariance condition implies that the MASAs generators should be all 
even or odd under the action of (\ref{ptgena}),  $Z_\mu=\varepsilon \mathcal{P}( \overline{Z}_\mu)\mathcal{P}$ 
where $\varepsilon^2=1$, in addition to the restrictions in Eqs. (\ref{transa}). Although the discussion so far was done at the classical level, the spirit of the non-Hermitian construction and the complex nature of the potentials is oriented mainly at the quantum level. Thus, once the classical reduction is done, the next step is to quantize the Hamiltonian, which will be performed in the simplest way possible. After finding the solutions of the Schr\"odinger equation, thanks to the integrability scheme, the focus is on the set of wave functions that are also compatible with the corresponding symmetry (\ref{ptgena}). In this form, the reality of the spectrum is guaranteed, unless there could exist regions of the parameters or couplings where the wave functions lose a fixed parity and the $\mathcal{PT}$-symmetry is spontaneously broken. The corresponding Schr\"odinger equations will be solved by means of separation of variables, which is related with the existence of second order invariants \cite{reviewsuper}. Before quantization, the corresponding conserved quantities can be studied in a similar way as the original scheme  \cite{InSys1,InSys3,InSys15,InSys2}. Once the reduced Hamiltonian system (real or complex) is defined, the integrals of motion and the Hamiltonian are the projections of the Lie algebra generators \cite{InSys1} as elements in the enveloping algebra. In other words, any generator $X$ of the Lie algebra associated to the symmetry group can be projected into a phase space function $X \rightarrow \widehat{X}$ by means of the momentum map \cite{Marsden,InSys1}
\begin{eqnarray}
\label{pGama}
\widehat{X}= p_q^T X q+p_r^T\,  \overline{X} \, r\,.
\end{eqnarray}
The mapping preserves the commutation relation at the level of Poisson brackets $\{\widehat{X}_i,\widehat{X}_j\}_{\text{PB}}=\widehat{X}([X_i,X_j])$ and can be used to project any function of the generators into the phase space. The application of this reduction scheme for the real and complex case yields
\begin{equation}
\label{redu}
\widehat{X}=\frac{1}{2}p_s^T(X+\overline{X})s+\frac{1}{2}k^T\mathcal{A}^{-1}(X-\overline{X})s \, ,
\qquad
\mathcal{A}_{\mu\nu}=(Z_\nu)_{\mu\sigma}s_\sigma\, .
\end{equation}
Since we are working with $U(n)$ where the Lie algebra 
associated is ${u}(1)\oplus {su}(n)$,  the generators $X$ are realized as anti-Hermitian  matrices and can be sort them as pure imaginary symmetric and real antisymmetric matrices \cite{InSys1}. In the current approach, the MASAs generators are chosen as linear combinations of symmetric matrices leading to,
 \begin{eqnarray}
\label{Genzenk}&
\widehat{Z}_\rho=
k^T \mathcal{A}^{-1} {Z}_{\rho}
 s= 
k_{\sigma} \mathcal{A}^{-1}_{\sigma\nu} \mathcal{A}_{\nu\rho}=k_\rho\,.&
\end{eqnarray}
In the same manner as the real case \cite{InSys1}, the integrals of motion of the system are obtained 
from the sub-space on the enveloping algebra of 
$su(n)$ which is invariant under transformations generated by the complex MASAs. 
These invariants 
are given as quadratic functions of the generators $X$, 
\begin{equation}
\label{IntTdef}
{T}_\ell=\sum_{i,j}c_{ij}^\ell{X}_{i}{X}_j\,,\qquad
[{T}_\ell,Z_\rho]=0\,,\qquad
c_{jk}^\ell \in \mathbb{C}\,,
\end{equation}
and they are projected into the reduced phase space via ${T}_\ell(\hat{X})\rightarrow \widehat{T}_\ell (\widehat{X})$. In order to illustrate how this scheme generates new $\mathcal{PT}$-invariant
super-integrable systems, the next sections focus in examples living in lower dimensional spheres.

\section{$\mathcal{PT}$-symmetric models on $S^1$}
\label{SecS1}
The simplest example can be studied is of a particle living in the unit circle $s_1^2+s_2^2=1$ which implies the constraint $s_1 p_1+s_2p_2=0$. We choose the $u(2)=u(1)\oplus su(2)$ algebra realized in terms of the Pauli matrices and commutation relations as follows 
\begin{equation}
\label{su(2)}
X_0= \Ti\sigma_0\,,\qquad 
X_1= \Ti\sigma_1\,,\qquad
X_2= \Ti\sigma_2\,,\qquad
X_3= \Ti\sigma_3\,, 
\end{equation}
\begin{equation}
\label{sua(2)}
 [X_3,X_2]=2X_1\,,\qquad
 [X_2,X_1]=2X_3\,,\qquad 
 [X_1,X_3]=2X_2\,.
\end{equation}
Clearly, only the $\{\sigma_0,\sigma_1,\sigma_3\}$ matrices are symmetric and compatible with the conditions for the MASAs in (\ref{transa}). Thus, without any loss of generality, the MASAs generators $Z_\mu$ can be chosen depending on two complex parameters $a$ and $b$, 
\begin{eqnarray}
\label{su2masas}
Z_1= X_0\,,\qquad
Z_2= a X_3-\Ti bX_1= \left(\begin{array}{cc}
\Ti a & b\\
b & -\Ti a
\end{array}\right)\,.
\end{eqnarray}  
By construction, the MASAs commute $[Z_1,Z_2]=0$ and the constants $a$ and $b$ are defined such that they cannot simultaneously vanish.
The real case with $a=1$ and $b=0$  has already been studied  
in \cite{InSys2}, yielding to the $n=2$ version of (\ref{example}). The MASAs and $X$'s generators in the reduced
phase space (\ref{redu}) take the form
\begin{equation}
\label{Zenk}
\widehat{Z}_1=\widehat{X}_0=k_1\,,\qquad \widehat{Z}_2=a \widehat{X}_3-\Ti b\widehat{X}_1=k_2\,,
\end{equation} 
\begin{equation}
\widehat{X}_1=\Ti \frac{a k_1+k_2 (s_2^2-s_1^2)}{b (s_1^2- s_2^2)-2\Ti  a s_1 s_2}\,,\qquad
\widehat{X}_2=s_1p_2-s_2p_1\,,\qquad
\widehat{X}_3=\frac{2 \Ti k_2 s_1 s_2- b k_1}{b (s_1^2- s_2^2)-2\Ti  a s_1 s_2}\, .
\end{equation}
It should be emphasised that the two parameters $a$ and $b$ are independent of the couplings $k_1$ and $k_2$, and in fact have a different basis. In this scheme, the couplings appear after the phase space reduction (\ref{Zenk}) from the vector $p_x=(k_1,k_2)$, unlike the parameters $a$ and $b$, which initially define the MASAs. The Casimir of the $su(2)$ algebra after the reduction is the Hamiltonian (\ref{genrealcase})  $\widehat{X}_1^2+\widehat{X}_2^2+\widehat{X}_3^2=2\widehat{H}_{a,b}$, where  
\begin{equation}
\widehat{H}_{a,b}=p_1^2+p_2^2+V_{a,b}(s_1,s_2)\,,
\end{equation}
and $V_{a,b}(s_1,s_2)$ represents the
two-parameter family of complex potentials (\ref{ComV}),
\begin{equation}\label{potone}
V_{a,b}(s_1,s_2)=\frac{2 k_1 k_2 (a (s_1^2-s_2^2)+2 \Ti b s_1 s_2)-k_1^2(a^2-b^2)-k_2^2 }{(b (s_1^2- s_2^2)-2\Ti  a s_1 s_2)^2}\,.
\end{equation}
Naturally, as the effective model in this case is one-dimensional, the Hamiltonian is the unique integral of motion. 
In order to discuss the quantum features of the Hamiltonian, setting $\hbar=1$ from now on, the polar coordinates  $(s_1,s_2)=(\cos\varphi,\sin\varphi)$ with  $\varphi \in [0,2\pi)$ are introduced  transforming the Hamiltonian into $\widehat{H}_{a,b}=p_\varphi^2+V_{a,b}(\varphi)$ with the potential   
\begin{equation}
\label{Potvarphi}
V_{a,b}(\varphi)=\frac{2 k_1 k_2 (a \cos 2 \varphi - \Ti b \sin 2 \varphi )-k_1^2 \left(a^2-b^2\right)-k_2^2}{(b \cos 2 \varphi -\Ti a \sin 2 \varphi )^2}\,.
\end{equation}
If the parameters $a$ and $b$ are real, the Hamiltonian displays an evident $\mathcal{PT}$-symmetry in terms of the usual parity and time-reversal operators
\begin{equation}\label{fpt}
\mathcal{P}: \varphi \rightarrow -\varphi \,, \qquad \mathcal{T}:
 \Ti\rightarrow-\Ti\,.
\end{equation}
Under particular values of the parameters, the potential $V_{a,b}(\varphi)$ reduces to several cases discussed in the literature, separated in three regimes $a=b$, $a>b$ and $b>a$. When $b=a$, (\ref{Potvarphi}) takes the form of the   
$\mathcal{PT}$-symmetric 
Morse potential, 
\begin{equation}\label{morse}
V_{a,a}(\phi)=\frac{k_2}{a^2}(2 a k_1   e^{-2 \Ti \varphi }-k_2 e^{-4 \Ti \varphi })\,.
\end{equation}
The change of variables $r= \Ti e^{\Ti \varphi}$ followed with
 $\psi(\varphi)=r^{-\frac{1}{2}}f(r)$ transmutes the Schr\"odinger equation $\widehat{H}_{a,b} \psi=E\psi$ into the equation $(-\sfrac{d^2}{dr^2}+\omega^2 r^2+g/r^2)f(r)={\cal E}f(r)$  with a radial harmonic oscillator potential \cite{Znojil1}. Here, the role of eigenvalues and couplings are swapped, $g=E-\frac{1}{4}$ and ${\cal E}=2k_1k_2/a$, in what is known as coupling constant metamorphosis and the St\"ackel transformation \cite{ccm}. In the remaining cases, the potential $V_{a,b}(\varphi)$ can be reduced to the P\"oschl-Teller type of potentials in terms of complex variables.  When $a>b$, the $\mathcal{PT}$-invariant variable $\xi$  defined by $\cos2\xi=\frac{a \cos2\varphi+\Ti b \sin 2\varphi}{\sqrt{a^2-b^2}}$ transforms (\ref{Potvarphi}) into 
\begin{equation}
\label{PTpot}
V_{a,b}(\xi)= \frac{g_-(g_{-}-1)}{\sin^2\xi}+\frac{g_+(g_{+}-1)}{\cos^2\xi}\,,
\end{equation}
where the new coupling constants $g_\pm$ are given in terms of the previous couplings and parameters $g_\pm(g_\pm{-}1)=\frac{(\sqrt{a^2-b^2}k_1\pm k_2)^2}{4(a^2-b^2)}$. The formal solutions of the Schr\"odinger equation with potential (\ref{PTpot}) can be written in terms of hypergeometric functions ${}_{2}F_{1}(a,b;c;z)$, 
\begin{align} \notag
\psi_n(\xi)&= 
c_1 (\sin \xi)^{g_-} (\cos \xi)^{g_+}  {}_{2}F_{1}\left(-n,g_-+g_++n;\sfrac{1}{2}+g_+;\cos^2\xi\right)\\ \label{wavef}
&+c_2 (\sin \xi)^{g_-} (\cos \xi)^{1-g_+}\,  {}_{2}F_{1}\left(\sfrac{1}{2}-n-g_+,\sfrac{1}{2}+n+g_-;\sfrac{3}{2}-g_+;\cos^2\xi\right)\, , 
\end{align}
with eigenvalues $E_n=(2n+g_-+g_+)^2$ and the integration constants $c_1$ and $c_2$. As usual, the quantum number $n$ must be defined taking into account the proper behaviour of the solutions $\psi_n(\xi)$. Among the different options that can be chosen, it is natural to require the wave functions to be single-valued, imposing that $g_\pm$ or $1-g_\pm$ must be integers, since they appear as powers of complex functions of the coordinate $\xi$. These restrictions imply that the quantum number $n$ must be integer or half-integer in the case $c_2=0$ or $c_1=0$, respectively, in order to have well-defined functions at $\pm \infty$. 
Furthermore, some of the singularities of the potential $V_{a,b}(\xi)$ and the wave functions $\psi_n(\xi)$ in the real line can be avoided because the variable $\xi$ is complex.
 However, the reality of the spectrum is ensured as long as the wave functions (\ref{wavef}) are also eigenstates of $\mathcal{PT}$ (\ref{fpt}) due to the invariance $\xi$, which is true for the above cases. 
Note that in this case the potential (\ref{PTpot}) can also be seen as the $\mathcal{PT}$-regularised 
trigonometric P\"oschl-Teller potential, obtained by applying the complex transformation 
$\varphi\rightarrow \varphi+\Ti \beta$ fixing 
$a=\cosh(2\beta)$ and $b=\sinh(2\beta)$ \cite{Znojil2, fcomikh}. The last case, where $b>a$, has to be treated separately because the couplings $g_\pm$ became complex, introducing multi-valued functions. Since the wave functions (\ref{wavef}) can also be expressed in terms of the Jacobi polynomials $P_n^{(\alpha,\beta)}(x)$, their complex extension and orthogonality can be studied by appropriate contours on a Riemann surface \cite{JacobiOr}. Since the focus here is on designing new models rather than exhausting all possible configurations, the analysis of the closed contours for $b>a$ is beyond the scope of this article.

\section{$\mathcal{PT}$-symmetric models on $S^{2}$}
\label{SecS2}
Real superintegrable systems in the two-dimensional sphere $s_1^2+s_2^2+s_3^2=1$ have been widely studied in the past \cite{contra, mil14,cal06,cal09, cal09v2, kmp, post11, vin19, kuru20, CorOlMar, thenew}, in particular, a complete analysis of the symmetry reduction method to obtain them was done in Ref. \cite{InSys2}. They studied the set of conserved quantities together with the separability of the Hamilton--Jacobi equation. Since the number of constraints on the allowed combinations of MASAs is higher in the real case, many of the possibilities were exhausted regarding the different systems that can be constructed using the $su(3)$ or $su(2,1)$ algebras. In contrast, for the non-Hermitian case, the possibilities to construct integrable models increase enormously due to the different MASAs together with their complex linear combinations that can be chosen.  The building blocks for the reduction scheme are the eight generators $X
_i$, $i=1,\ldots,8$ of the $u(3)=u(1)\oplus su(3)$ algebra
\begin{eqnarray}
\label{su(3)}
&\begin{array}{lll}

[X_1,X_3]=2X_4\,,&
[X_1,X_4]=-2X_3\,,& 
[X_1,X_5]=X_6\,,
\\

[X_1,X_6]=-X_5\,,&
[X_1,X_7]=-X_8\,,&
[X_1,X_8]=X_7\,,\\

[X_2,X_3]=-X_4\,,&
[X_2,X_4]=X_3\,,&
[X_2,X_5]=X_6
\,, \\

[X_2,X_6]=-X_5\,,&
[X_2,X_7]=2X_8\,,&
[X_2,X_8]=-2X_7\,,\\

[X_3,X_4]=2X_1\,,&
[X_3,X_5]=- X_7\,,&
[X_3,X_6]=- X_8\,,\\

[X_3,X_7]=X_5\,,&
[X_3,X_8]=X_6\,,&
[X_4,X_5]= X_8\,,\\

[X_4,X_6]=- X_7\,,&
[X_4,X_7]=X_6\,,&
[X_4,X_8]=-X_5\,,\\

[X_5,X_6]=2(X_1+X_2)\,,&
[X_5,X_7]=-X_3\,,&
[X_5,X_8]= X_4\,,\\

[X_6,X_7]= -X_4\,,&
[X_6,X_8]= -X_3\,,&
[X_7,X_8]=2X_2\,.
\end{array}
\end{eqnarray} 
Since the symmetry structure of the mechanical systems arise from the $su(3)$ enveloping algebra  (\ref{su(3)}), the quadratic and cubic Casimir elements,
\begin{align}
{\cal C}_{2}&= 4 (X_1^{2}+X_2 X_1+ X_{2}^{2}) +3(X_3^{2} + X_4^{2} + X_5^{2} +X_6^{2} +X_7^{2}+ X_8^{2})\, , \\ \label{cas3}
{\cal C}_{3}&=(X_8 X_6{+}X_7 X_5) X_4{+}(X_8 X_5{-} X_7 X_6) X_3{+}\sfrac{4}{27}\left(X_1{-}X_2\right)\left(2X_1{+}X_2\right)\left(X_1{+}2X_2\right)+\\ \notag
&+\sfrac{1}{6}\{X_1{+}2X_2,X_3^2{+}X_4^2\}{+}\sfrac{1}{6}\{X_1{-}X_2,X_5^2{+}X_6^2\}{-}\sfrac{1}{6}\{2X_1{+}X_2,X_7^2{+}X_8^2\}{-}\sfrac{4}{3}(X_1{-}X_2)\, ,
\end{align}
where $\{A,B\}=AB+BA$, are useful to understand the features of the polynomial algebra \cite{CorOlMar}. In particular, the Hamiltonian is usually identified in terms of the quadratic Casimir operator ${\cal C}_{2}$ which, at the same time, can be expressed as combinations of the integrals of motions. The algebra (\ref{su(3)}) can be realized as $3\cross 3$ anti-Hermitian matrices
\begin{align}
\label{X1}
X_1&=
\left(\begin{array}{ccc}
\Ti &0 &0 \\
0 & -\Ti & 0\\
0 & 0 & 0
\end{array}\right)\,,&
X_2&=
\left(\begin{array}{ccc}
0 &0 &0 \\
0 & \Ti & 0\\
0 & 0 & -\Ti
\end{array}\right)\,,&
X_3&=
\left(\begin{array}{ccc}
0 &1 &0 \\
-1 & 0 & 0\\
0 & 0 & 0
\end{array}\right)\,,& \\ \label{X4}
X_4&=
\left(\begin{array}{ccc}
0 &\Ti &0 \\
\Ti & 0 & 0\\
0 & 0 & 0
\end{array}\right)\,,&
X_5&=
\left(\begin{array}{ccc}
0 &0 &1\\
0 & 0 & 0\\
-1 & 0 & 0
\end{array}\right)\,,&
X_6&=
\left(\begin{array}{ccc}
0 &0 &\Ti \\
0 & 0 & 0\\
\Ti & 0 & 0
\end{array}\right)\,,&
\\
X_7&=
\left(\begin{array}{ccc}
0 &0 &0 \\
0 & 0 & 1\\
0 & -1 & 0
\end{array}\right)\,,&
X_8&=
\left(\begin{array}{ccc}
0 &0 &0 \\
0 & 0 & \Ti\\
0 & \Ti & 0
\end{array}\right)
\,,&
X_{0}&=\Ti \I_{3\cross 3}\,,\label{X8}&
\end{align}
where only the matrices $\{X_0,X_1,X_2,X_4,X_6,X_8\}$ are symmetric. The non-Hermitian generalisation allows to take any commuting complex linear combination of the symmetric set as MASAs generators $Z_\mu$ of (\ref{transa}), extending the class of systems with respect to those not admissible in the real case \cite{InSys1,InSys2}.  After the reduction to the phase space using Eq. (\ref{redu}), the remaining antisymmetric generators are mapped into the angular momentum components 
\begin{equation}\label{gro}
\widehat{X}_{7}=-L_1\,,\qquad
\widehat{X}_{5}=L_2\,,\qquad
\widehat{X}_{3}=-L_3\,,\qquad
L_i=\epsilon_{ijk}s_j p_{k}\, .
\end{equation}
Since the dynamics lives in the $S^{2}$ surface, imposing the constraint $s_1p_1+s_2p_2+s_3p_3=0$, the sum of the angular momentum squared reduces to the usual kinetic term as canonical momentum squared 
\begin{equation}\notag
L^2=L_1^2+L_2^2+L_3^2|_{S^2}=p_1^2+p_2^2+p_3^2 \, .
\end{equation}

\subsection{One parameter complex 2D integrable model} \label{4p1}
As already pointed out, the main difference between the real standard method \cite{InSys1,InSys3,InSys15,InSys2} and the non-Hermitian approach presented here is the relaxed choice of the $Z_\mu$ matrices as complex linear combinations of the symmetric subset of $su(3)$ generators (\ref{X1})-(\ref{X8}). 
In the same spirit as in the real case, the linear combinations are then uniquely determined by the condition $[Z_\mu, Z_\nu]=0$, with the freedom of the complex linear combinations allowing a large set of free parameters to design $\mathcal{PT}$-invariant Hamiltonians. Besides the couplings $k_1,k_2$ and $k_3$ coming from the reductions $Z_\mu \rightarrow \widehat{Z}_\mu=k_\mu$ via Eq. (\ref{Genzenk}), additional parameters can be introduced from the combinations of the allowed symmetric (\ref{transa}) MASAs $\{X_0,X_1,X_2,X_4,X_6,X_8\}$.  Since the motivation is to construct quantum models with real energies that exhibit features of non-Hermitian systems, such as spontaneous $\mathcal{PT}$ symmetry breaking, it is sufficient to demonstrate the mechanism with systems having only one extra parameter, namely $\lambda$, in addition of the three couplings $k_\mu$. Thus, an interesting choice of MASAs generators $[Z_1,Z_2]=[Z_2,Z_3]=[Z_3,Z_1]=0$ is given by 
\begin{align}\notag
    Z_1&= 
\frac{1}{3 \sqrt{1-2 \lambda ^2}}\left[(1-2 \lambda^2) X_0+ \gamma_- X_1+
   (1+\lambda^2)X_2-\frac{3}{2} \lambda^2 X_4+3\Ti \lambda(\lambda_- X_6
  -\lambda_+X_8) \right]\,, \\ \notag
      Z_2&= 
\frac{1}{3 \sqrt{1-2 \lambda ^2}}\left[(1-2 \lambda^2) X_0+ \gamma_+ X_1+
   (1+\lambda^2)X_2-\frac{3}{2} \lambda^2 X_4+3\Ti \lambda(\lambda_+ X_6
  -\lambda_-X_8) \right]\,, \\
      Z_3&= 
  \frac{1}{3 \sqrt{1-2 \lambda ^2}}\left[-(1-2 \lambda ^2) X_0+(1+\lambda^2)(X_1+2X_2)
  -3 \lambda^2 X_4+3 
  \Ti\lambda(X_6-X_8)\right]\,,
  \label{zmat}
\end{align}
that depend on a single parameter $\lambda$ that it ranges as $0<\lambda^2<\frac{1}{2}$ and defines the constants $\lambda_\pm$ and $\gamma_\pm$,
\begin{equation}\notag
\lambda_\pm=\frac{1}{2} \left(1\pm\sqrt{1-2 \lambda ^2}\right) , \qquad   \gamma_\pm=\frac{1}{2}(1+\lambda ^2 \pm 3 \sqrt{1-2 \lambda ^2}) \,. 
 \end{equation}
The reduction to the phase space formulation transforms the MASAs into the coupling constants $\widehat{Z}_\mu=k_\mu$ (\ref{Genzenk}) while the quantized version of the Hamiltonian computed with Eqs. (\ref{genrealcase}) and  (\ref{ComV}) takes the following form,
 \begin{equation}\label{hlam}
 \widehat{H}_\lambda=p_1^2+p_2^2+p_3^2+V_\lambda(s_1,s_2,s_3) =L_1^2+L_2^2+L_3^2+V_\lambda(s_1,s_2,s_3) \,, \qquad
 \end{equation}
 \begin{equation}\label{potm}
V_\lambda(s_1,s_2,s_3)=\frac{k_1^2}{\left(\lambda_- s_1-\lambda_+ s_2+\Ti \lambda  s_3\right)^2}+
\frac{k_2^2}{\left(\lambda_+ s_1-\lambda_- s_2+\Ti \lambda  s_3\right){}^2}+\frac{k_3^2}{\left(\Ti \lambda  (s_1-s_2)-s_3\right){}^2} \,.
 \end{equation}
 The Hamiltonian, up to some additive and multiplicative constants, coincides with the projection of the second order $su(3)$ Casimir  ${\cal C}_2$ (\ref{cas3}). Since the constants $\lambda_\pm$ are real in the range $0<\lambda^2<\frac{1}{2}$ the potential  $V_\lambda(s_1,s_2,s_3)$ is clearly complex and therefore the Hamiltonian is non-Hermitian. Defining the anti-linear symmetry in terms of a composed parity inversion in the $s_3$ coordinate and the usual conjugate operator,
 \begin{equation}
\label{PTtra}
\mathcal{P}: s_3\rightarrow -s_3 
\qquad \mathcal{T}: \Ti\rightarrow-\Ti\,,
\end{equation}
the potential (\ref{potm}) remains invariant under their combined action
  \begin{equation}
\mathcal{PT}\,V_\lambda(s_1,s_2,s_3)=\overline{V_\lambda}(s_1,s_2,-s_3)=V_\lambda(s_1,s_2,s_3) \, .
\end{equation}
In this way, following the ideas from non-Hermitian theories \cite{ali, znojilconf, rev}, it is quite suggestive that the Hamiltonian (\ref{hlam}) displays a real spectrum, if their wave functions are also eigenstates of the operator (\ref{PTtra}). Before concentrating on the eigenvalue problem, some remarks about the integrals of motion and the separability of the quantum problem are in order. In fact, the corresponding stationary Schr\"odinger equation $ \widehat{H}_\lambda\psi=E\psi$ is not separable in Cartesian coordinates, but the existence of a separable coordinate system is ensured by the 
presence
 of quadratic integrals of motion \cite{reviewsuper, Miller}. Since the system (\ref{hlam}) is effectively two-dimensional because of the spherical constraint, only $2\times2-1=3$ independent integrals of motion are expected, where two of them are mutually commuting. In fact, an over-complete set of integrals of motion $[\widehat{H}_\lambda, \widehat{T}_\mu]=0$, $\mu=1,2,3$ can be obtained as quadratic functions of the form (\ref{IntTdef}), 
  \begin{align}
 {T}_1&=(\lambda_- {X}_7{+}\lambda_+ {X}_5{+}\Ti \lambda  {X}_3)^2-\frac{[2 \lambda(\lambda _+ {X}_1{+} {X}_2){-}\lambda {X}_4{+} \Ti(\lambda ^2{+}\lambda _+) {X}_6{-} \Ti(\lambda ^2{+}\lambda _-) {X}_8]^2}{1-2\lambda^2}{+}4{Z}_2{Z}_3 \, ,\\
 {T}_2&=( \lambda_+ {X_7}{+}\lambda_- {X}_5{+}\Ti \lambda {X}_3)^2-
 \frac{[2\lambda(\lambda_-{X}_1{+}{X}_2){-}\lambda {X}_4{+}\Ti(\lambda^2{+}\lambda_-){X}_6{-}
 \Ti(\lambda^2{+}\lambda_+){X}_8]^2}{1-2\lambda^2}{+}4{Z}_1{Z}_3 \, ,\\
 {T}_3&= (\Ti \lambda  {X}_7{+}\Ti \lambda  {X}_2{-}{X}_3)^2+
 \frac{[\lambda ^2 {X}_1+2 \lambda ^2 {X}_2+\left(\lambda ^2-1\right) {X}_4+ \Ti\lambda  {X}_6-\Ti \lambda  {X}_8]^2}{1-2 \lambda ^2}\, .
 \end{align}
 Their reductions to the phase space are 
 \begin{align}\label{t1}
 \widehat{T}_1&=\left(\lambda_- L_1-\lambda_+ L_2+\Ti \lambda  L_3\right)^2+\left(k_2
 \frac{\Ti \lambda  s_1-\Ti \lambda  s_2-s_3}{\lambda_+ s_1-\lambda_- s_2+\Ti \lambda  s_3}+k_3 \frac{\lambda_+ s_1-\lambda_- s_2+\Ti \lambda  s_3}{\Ti \lambda  s_1-\Ti \lambda  s_2-s_3}
 \right)^2\, ,\\ \label{t2}
  \widehat{T}_2&=\left( \lambda_+ L_1-\lambda_- L_2+\Ti \lambda L_3   \right)^2+\left(k_1
 \frac{\Ti \lambda  s_1-\Ti \lambda  s_2-s_3}{\lambda_- s_1-\lambda_+ s_2+\Ti \lambda  s_3}+k_3\frac{\lambda_- s_1-\lambda_+ s_2+\Ti \lambda  s_3}{\Ti \lambda  s_1-\Ti \lambda  s_2-s_3} 
 \right)^2\, ,\\ \label{t3}
  \widehat{T}_3&=\left( \Ti \lambda  L_1-\Ti \lambda  L_2-L_3 \right)^2+\left(
k_1 \frac{\lambda_+ s_1-\lambda_- s_2+\Ti \lambda  s_3}{\lambda_- s_1-\lambda_+ s_2+\Ti \lambda  s_3}+k_2\frac{\lambda_- s_1-\lambda_+ s_2+\Ti \lambda  s_3}{\lambda_+ s_1-\lambda_- s_2+\Ti \lambda  s_3}
 \right)^2\, .
\end{align}
This set of integrals and the Hamiltonian (\ref{hlam}), therefore the Casimir ${\cal C}_2$ as well, are related through the relation, 
\begin{equation}
 \widehat{T}_1+\widehat{T}_2+\widehat{T}_3=(1-2\lambda^2)\widehat{H}_\lambda-(k_1-k_2-k_3)^2\,,
\end{equation}
establishing the superintegrability of the model as there are only three independent integrals. The structure of the integrals  (\ref{t1}), (\ref{t2}) and (\ref{t3}) is reminiscent of the well-known model on the 2-sphere with the Hamiltonian (\ref{example}) and the symmetry algebra turns out to be of the same form. In this case, $ \widehat{T}_1$, $\widehat{T}_2$ and $\widehat{T}_3$ constitute a non-Hermitian realization of the Racah algebra $R(3)$ \cite{kuru20,CorOlMar}. The structure of the algebra, which is not completely relevant to be included here, contains the elements 
$ \widehat{T}_{ij}=-\Ti[\widehat{T}_i, \widehat{T}_j ]$ which are also related with the second and third order Casimir \cite{CorOlMar}. These generators are not independent $\widehat{T}_{12}=-\widehat{T}_{13}=\widehat{T}_{23}$ and satisfy an algebra of the form
\begin{align}\label{t12}
[\widehat{T}_{12}, \widehat{T}_1 ]\propto \Ti\{\widehat{T}_1, \widehat{T}_3\}-\Ti\{\widehat{T}_1,\widehat{T}_2\}+\text{lower terms}\, , \\ \label{t12a}
[\widehat{T}_{12}, \widehat{T}_2 ]\propto \Ti\{\widehat{T}_1, \widehat{T}_2\}-\Ti\{\widehat{T}_2,\widehat{T}_3\}+\text{lower terms}\, .
\end{align}
where $\widehat{T}_{12}^2=P(\widehat{T}_1,\widehat{T}_2,\widehat{T}_3)$ is a polynomial in the integrals. The eigenvalue problem can be addressed using separation of variables of the Schr\"odinger equation by means of the coordinates $\xi$
and $\chi$, defined by the equations   
 \begin{equation}\label{coor}
 \cos 2\xi =\frac{(\lambda_- s_1-\lambda_+ s_2+\Ti \lambda s_3)^2-(\lambda_+ s_1-\lambda_- s_2+\Ti \lambda s_3)^2}{(\lambda_- s_1-\lambda_+ s_2+\Ti \lambda s_3)^2+(\lambda_+ s_1-\lambda_- s_2+\Ti \lambda s_3)^2}\,,\quad
 \cos \chi =\frac{\Ti\lambda(s_1-s_2)-s_3}{\sqrt{1-2\lambda^2}}\,,
 \end{equation}
being both $\xi$ and $\chi$ invariants under the  $\mathcal{PT}$-symmetry defined in Eqs. (\ref{PTtra}).
For $\lambda^2\neq \frac{1}{2}$, the transformation $(z_1,z_2,z_3)=(\cos\xi\sin\chi,\sin\xi\sin\chi,\cos\chi)$ with $z_\mu z^{\mu}=1$ allows to write $\widehat{H}_\lambda$ as the the generic model on $S^2$ (\ref{example}). However, here, the value of $\lambda$ intrinsically controls the 
$\mathcal{PT}$-symmetric nature of the system since the spectrum, coordinates and wave functions 
will depend on it. Explicitly, the equation $ \widehat{H}_\lambda\psi=E\psi$ with the uniparametric potential $V_\lambda(s_1,s_2,s_3)$ takes the form 
  \begin{equation}\label{Schr3}
-\left(
\frac{\partial^2\psi}{\partial\chi^2}+\cot \chi \frac{\partial\psi}{\partial\chi}
+\frac{1}{\text{sin}^2\chi}\frac{\partial^2\psi}{\partial \xi^2}\right)+\frac{1}{1-2\lambda^2}\left[
\frac{1}{\text{sin}^2\chi}\left(
\frac{k_1^2}{\text{cos}^2\xi}+\frac{k_2^2}{\text{sin}^2\xi}
\right)+\frac{k_3^2}{\cos^2\chi}\right]\psi=E\psi \, 
 \end{equation}
The separation of variables is then performed via $\psi(\chi,\xi)=\Psi(\chi)\Phi(\xi)$, where the coupling constants $k_\mu$ are reparametrised by means of  $\ell_\mu=\sfrac{1}{2}\left(1{+}\sqrt{1{+}\frac{4k_\mu^{2} }{1{-}2\lambda^2}}\right)$ with  $\mu=1,2,3$,
\begin{align}\label{sep1}
-\frac{\partial^2\Phi(\xi)}{\partial \xi^2}+\left(
\frac{\ell_1(\ell_1-1)}{\text{cos}^2\xi}+\frac{\ell_2(\ell_2-1)}{\text{sin}^2\xi}
\right)\Phi(\xi)=(\ell_1+ \ell_2 + 2m)^2\Phi(\xi)\,, \\ \label{sep2}
-
\frac{\partial^2\Psi(\chi)}{\partial \chi^2}-\cot\chi\frac{\partial\Psi(\chi)}{\partial\chi}
+ \left(\frac{\ell_3(\ell_3-1)}{\text{cos}^2\chi}+\frac{(\ell_1+ \ell_2 + 2m)^2}{\text{sin}^2\chi}\right)\Psi(\chi)=E\Psi(\chi)\,.
\end{align}
The constant $m$ plays the role of quantum number and was introduced as separation constant between two quantum systems with P\"oschl-Teller potentials in complex coordinates. As in the previous section, the solutions of the Eqs. (\ref{sep1}) and (\ref{sep2}) can be expressed in terms of the hypergeometric functions and the integration constants $c_i$, $i=1,...,4$
\begin{align} \notag
\Phi_m(\xi)&= 
c_1 (\sin \xi)^{\ell_{2}} (\cos \xi)^{\ell_{1}}\,  {}_{2}F_{1}\left(-m,\ell_1+\ell_2+m;\sfrac{1}{2}+\ell_1;\cos^2\xi\right)\\
&+c_2 (\sin \xi)^{\ell_{2}} (\cos \xi)^{1-\ell_{1}}\,  {}_{2}F_{1}\left(\sfrac{1}{2}-m-\ell_1,\sfrac{1}{2}+m+\ell_2;\sfrac{3}{2}-\ell_1;\cos^2\xi\right)\, \label{eigen1} ,\\ \notag
\Psi_{m,n}(\chi)&= c_3 (\sin \chi)^{-\ell_{1}-\ell_{2}-2m} (\cos \chi)^{\ell_{3}}\,  {}_{2}F_{1}\left(-n,\sfrac{1}{2}{-}\ell_1{-}\ell_2{+}\ell_3{-}2m{+}n;\sfrac{1}{2}+\ell_3;\cos^2\chi\right)\\
&+c_4 (\sin \chi)^{-\ell_{1}-\ell_{2}-2m} (\cos \chi)^{1-\ell_{3}}\,  {}_{2}F_{1}\left(\sfrac{1}{2}-n-\ell_3,
1{-}\ell_1{-}\ell_2{-}
2m{+}n
;\sfrac{3}{2}-\ell_3;\cos^2\chi\right)\, . \label{eigen2} 
\end{align}
The solutions $\Psi_{m,n}(\chi)$ depend on the extra quantum number $n$, which must be specified along with the solutions of interest. The avoidance of multi-valued functions imposes constraints on the powers of the trigonometric functions in (\ref{eigen1}) and (\ref{eigen2}) to have integer values, namely $\ell_1$, $\ell_2$, $\ell_3$, $1-\ell_1$, $-\ell_{1}-\ell_{2}-2m$ and so on. Of course, these choices must simultaneously be compatible with the regularity of the solutions in the asymptotic regime, which induces additional conditions on the first argument in the hypergeometric functions and on the quantum numbers $n$ and $m$. In this large landscape of possibilities, both wave functions show a well-defined parity under (\ref{PTtra}) as long as the parameter $\lambda$ is between $0<\lambda^2<\frac{1}{2}$,
\begin{equation}
\mathcal{PT}\, \Phi_m(\xi) = \varepsilon\, \Phi_m(\xi)\, , \quad \mathcal{PT}\, \Psi_{m,n}(\chi) = \varepsilon\, \Psi_{m,n}(\chi) ,\quad \varepsilon^2=1 \, .
\end{equation}
In this way, the energies of both wave functions in (\ref{eigen1}) and (\ref{eigen2}) are real, defined in terms of the couplings and quantum numbers, 
\begin{equation}
E_{n,m}=(\ell_1+ \ell_2-\ell_3+2m-2n-1)
(\ell_1+\ell_2-\ell_3 +2m -2n) \, .
\end{equation}
However, if the parameter moves away from the valid region, i.e. $\lambda^2>\frac{1}{2}$, the system enters a $\mathcal{PT}$-broken phase. The coordinates $\xi$ no longer have a definite parity under the $\mathcal{PT}$ transformation, the potentials in (\ref{Schr3}) are repulsive\footnote{In the real case, some of these regions of parameters have to be treated from the point of view of self-adjoint extensions \cite{self}, so it is expected that a rigorous analysis should be considered in the complex case.} and eventually the energies become complex, together with the couplings $\ell_\mu$. Interestingly, in $\lambda^2>\frac{1}{2}$ a new type of anti-linear symmetry arises, with the same $\mathcal{T}$ operator as in (\ref{PTtra}), but defining a new parity operator $\widetilde{\mathcal{P}}$ as follows,
\begin{equation}
\label{PTtra2}
\widetilde{\mathcal{P}}: s_1\leftrightarrow s_2 \, .
\end{equation}
In this case, a completely new analysis has to be performed taking into account the complex nature of the couplings $\ell_\mu$ and the complex analysis of the hypergeometric functions, in a similar fashion to the one dimensional case in the previous section for $b>a$.  Finally, in the particular case $\lambda^2=\frac{1}{2}$ the matrices 
$Z_\mu$ in Eq. (\ref{zmat}) are ill-defined, but they can be rescaled in such a way that the reduced Hamiltonian (\ref{genrealcase}) with the potential (\ref{ComV}) can be computed, 
\begin{equation}\label{last}
 \widehat{H}_{\pm\frac{1}{\sqrt{2}}}=p_1^2+p_2^2+p_3^2+\frac{\alpha^2}{(s_1-s_2\pm \Ti \sqrt{2} s_3)^2}\, ,
\end{equation}
where the potential is now invariant with respect to both transformations $\mathcal{PT}$ and $\widetilde{\mathcal{P}}\mathcal{T}$ and $\alpha^2=4 k_1^2+4 k_2^2-2 k_3^2$ is just a redefinition of the coupling constants. In this case the three potential terms in (\ref{potm}) collapse into just one as well as 
the three integrals of motion (\ref{t1})-(\ref{t3}) are reduced to $\widehat{T}=L_1-L_2\pm \Ti \sqrt{2}L_3$, $[\widehat{T},  \widehat{H}_{\pm\frac{1}{\sqrt{2}}}]=0$. The superintegrability of the system (\ref{last}) can be understood in terms of a new variable $z=(s_1-s_2\pm \Ti \sqrt{2} s_3)^{-1}$ such that the effective Schr\"odinger equation $ \widehat{H}_{\pm\frac{1}{\sqrt{2}}}\psi=E\psi$ takes the form,
\begin{equation}\label{new2}
\left(-\frac{d^{2}}{dz^2}+\frac{E}{z^2}\right)\psi(z)=\alpha^2\psi(z)\,.
\end{equation}
which is another case of the coupling constant metamorphosis \cite{ccm}, where the roles of the energy and the coupling constant are exchanged. The solutions of (\ref{new2}) are given in terms of Bessel functions, expanded as 
\begin{equation}
\psi_{\alpha,q}(z)=\sum_{j=0}^{\infty}
\frac{(-1)^{j}}{j!\Gamma(j+q+\frac{3}{2})}\left(\frac{\alpha z}{2}\right)^{2j+q+1}
\end{equation}
where $E=q (q+1)$ and $\Gamma(x)$ is the gamma function. Single-valued functions of $\psi_{\alpha,q}(z)$ appear when $\ell$ takes integer values being at the same time eigenstates of both $\mathcal{PT}$ and $\widetilde{\mathcal{P}}\mathcal{T}$ operators. Hence, the energies $E$ of the Hamiltonian (\ref{last}) are real. To conclude this subsection, it is worth looking at the results obtained from another point of view. 
Superintegrable systems on the 2-sphere with second order invariants are separated in five coordinate systems \cite{Miller, kalnins, recent}, including spherical ones. 
In this sense, it is expected that the MASAs approach will find cases that are related to the systems classified in the Refs. \cite{comp2, comp3, comp4}. What is surprising, however, is that by increasing the number of free parameters in the MASAs, one can also generate new superintegrable non-Hermitian systems, which contain models in these classifications as special cases. In addition, the parameters in this description can determine the physical nature of the systems in a slightly different way than in the real case.  The potentials (\ref{potone}) and (\ref{potm}) are manifest examples of these features in one- and two- dimensions since they relate to the  generic model in the sphere through a complex transformation, that is valid only for a restricted range of the parameters. Such transformations do not simply eliminate the parameters as they appear explicitly in the spectrum and the wave functions characterising the 
 $\mathcal{PT}$-symmetric phase of the systems. More of these cases will be shown in section \ref{4p2} below.

\subsection{More non-Hermitian superintegrable models}\label{4p2}

The non-Hermitian scenario for the Marsden--Weinstein mechanism allows the generation of new complex potentials that, for certain values of the parameters, are reduced or related to known results in the literature. On the one hand, the reduction scheme on the free particle in the hyperbolic background $s_1^2+s_2^2-s_3^2=1$ \cite{InSys2, boyer}, produces Hamiltonians with Minkowski metric signature in the kinetic terms. The reason behind for this kind of kinetic terms in the real case are the different classes of MASAs $su(2,1)$ subalgebras, known as the non-compact Cartan subalgebra, the orthogonally decomposable subalgebra, and the nilpotent subalgebra  \cite{InSys1,InSys2}.  Considering that the physical interpretation of such models is more elusive, in the non-Hermitian approach these superintegrable systems appear with the Euclidean kinetic term and as particular cases of a more general multiparameter family of systems.  On the other hand, a number of papers have addressed the classification of separable coordinate systems in the complex two-sphere  \cite{comp2, comp3, comp4}, leading to complex potentials with three independent couplings and some other particular cases. In this context, one of the advantages of the present approach is that it is possible to generalise and include some of these superintegrable systems with suitable linear combination MASAs and free parameters. This can be seen in a simple example where the MASAs are defined according to two parameters $a$ and $b$,
\begin{equation}\label{masas1}
{Z}_1=\frac{1}{3} ({X}_0+2 {X}_1+X_2)\,,\quad
{Z}_2=a {X}_2+2 \Ti b {X}_8\,, \quad
{Z}_3=\frac{1}{3} (2 {X}_0-2 {X}_1-{X}_2)\, .
\end{equation}
The logic underlying the construction is the same as that presented above, repeating the steps to construct the potential, the Hamiltonian and the conserved quantities. From (\ref{genrealcase}) and (\ref{ComV}) the following family of potentials is obtained,
\begin{equation}\label{spe1}
V_{a,b}=
\frac{k_1^2}{s_1^2} +\frac{\left(s_2^2+s_3^2\right) \left(k_2^2+\left(a^2-4 b^2\right) k_3^2\right)-2 k_2 k_3 \left(a (s_2^2-s_3^2)+4 b \Ti s_3 s_2\right)}{4 \left(a s_2 s_3-\Ti b \left(s_2^2-s_3^2\right)\right){}^2}
\, .
\end{equation}
The set of quadratic generators (\ref{IntTdef}) commuting with the complex MASAs (\ref{masas1}) can be chosen as
\begin{align}
{T}_{1}&={X}_3^2+{X}_4^2+{X}_5^2+{X}_6^2\, , \\
{T}_{2}& ={X}_2^2+{X}_7^2+{X}_8^2\,,\\
{T}_{3}& =a ({X}_3^2+{X}_4^2)+\Ti b(\{X_3,X_5\}+\{X_4,X_6\})\,,
\end{align}
and their reduction to the phase space via (\ref{redu}) yields
\begin{align}
\widehat{T}_{1}&=L_2^2+L_3^2+\frac{k_1^2}{s_1^2} +s_1^2V_{a,b} +2(k_3k_1-k_1^2) \, ,\\
\widehat{T}_{2}& =L_1^2+(1-s_1^2)\left(V_{a,b} -\frac{k_1^2}{s_1^2} \right) \, ,\\
\widehat{T}_{3}&=a L_3^2 +\Ti b \{L_2 ,L_3\}+s_1^2\frac{(k_2 s_3+k_3[as_3-2\Ti b s_2])([a^2-4b^2]k_3 s_3+k_2[as_3-2\Ti b s_2])}{4 \left(a s_2 s_3-\Ti b \left(s_2^2-s_3^2\right)\right){}^2}) \\
&+\frac{k_1^2}{s_1^2} s_2 (a s_2{+}2 \Ti b s_3)+k_1(k_2+a k_3) \,.
\end{align}
The Hamiltonian defined by $ \widehat{H}_{a,b}=p_1^2+p_2^2+p_3^2+V_{a,b}$ and the 
integrals $[\widehat{H}_{a,b},\widehat{T}_i]=0$, $i=1,2,3$ have several similarities with the two-dimensional case in section \ref{4p1}. All conserved quantities are invariant under the same $\mathcal{PT}$-symmetry defined above (\ref{PTtra}) and the integrals are an over-complete set satisfying 
$\widehat{T}_{1}+\widehat{T}_{2}=\widehat{H}_{a,b} +2k_1k_3-k_1^2$. The last relation indicates that the quadratic algebra of the integrals will be slightly different than (\ref{t12}) and (\ref{t12a}) considering again $ \widehat{T}_{12}=-\Ti[\widehat{T}_1, \widehat{T}_2 ]$. The nature of the algebra can also be understood by taking into account some of the limit cases of the potential (\ref{spe1}). The two particular cases $b=1/2$, $a=0$ and $b=1/2$, $a=1$ are the non-Hermitian counterparts of the non-compact Cartan and orthogonally decomposable subalgebra MASAs reductions \cite{InSys1,InSys2} respectively,  
\begin{align}
V_{0,\frac{1}{2}}&=
\frac{k_1^2}{s_1^2}+\frac{(k_3^2-k_2^2) (s_2^2+s_3^2)+4 \Ti k_2 k_3 s_2 s_3}{\left(s_2^2-s_3^2\right){}^2}=\frac{k_1^2}{s_1^2}-\frac{(k_2-\Ti k_3)^2}{2(s_2-s_3)^2}-\frac{(k_2+\Ti k_3)^2}{2(s_2+s_3)^2}\,, \\
V_{1, \frac{1}{2}}&=
\frac{k_1^2}{s_1^2}+\frac{2 k_2 k_3}{\left(s_2+\Ti s_3\right){}^2}
-\frac{k_2^2 \left(s_2-\Ti s_3\right)}{\left(s_2+\Ti s_3\right){}^3}\,.
\end{align}
The first potential can be understood as a rotation in the $s_2-s_3$ plane followed by a complex redefinition of the couplings from the generic model on the sphere (\ref{example}). The second one appears in the classification scheme of superintegrable Hamiltonians in the complex two-sphere \cite{comp2}. The type of separable coordinate systems is related to the superintegrable systems and their particular algebra and the form of $ \widehat{T}_{12}^2$ \cite{recent}, which allows the classification of these systems. For this reason, it is expected to find different algebraic structures among the possible complex MASAs configurations. In the same way, the analogue of the nilpotent subalgebra case can be designed with the following MASAs in terms of nilpotent generators ${Z}_2^2={Z}_3^3=0$,
\begin{equation}
{Z}_1= {X}_0\,,\qquad
{Z}_2=  {X}_2+\Ti {X}_8\,,\qquad
{Z}_3= {X}_4+\Ti {X}_6\,.
\end{equation}
The complex potential with this choice takes a similar form as in Refs. \cite{InSys2,comp2}, 
\begin{equation}
V_{\text{N}}=\frac{2 k_1 k_2+k_3^2}{\left(s_2+\Ti s_3\right){}^2}-\frac{4 k_2 k_3 s_1}{\left(s_2+\Ti s_3\right){}^3}
+\frac{k_2^2 \left(4 s_1^2-1\right)}{\left(s_2+\Ti s_3\right){}^4} \, ,
\end{equation}
and shows the same $\mathcal{PT}$-symmetry in the previous examples. The integrals of motion are also constructed from the reduction of the invariant elements in the enveloping algebra, namely,
\begin{align}
{T}_{1}&=\frac{4}{3}X_1^2+\frac{2}{3}X_2^2+2X_3^2+X_4^2+X_6^2+X_7^2+\sfrac{1}{3}X_8^2
+\Ti \{X_3,X_5\}-\frac{2}{3}\{X_1,X_2+2\Ti X_8\} \, ,\\
{T}_{2}&=-(X_3+\Ti X_5)^2+\sfrac{2}{3}\{2X_1+X_2,X_2+\Ti X_8\} \, , \\
{T}_{3}&=-\sfrac{1}{6}\{2X_1+X_2,X_4\}-\sfrac{\Ti}{6}\{2X_1-5X_2-6\Ti X_8,X_6\}+\sfrac{1}{2}\{\Ti X_3-X_5,X_7\} \, .
\end{align}
Their invariants also form an over-complete set that can be written in terms of the Hamiltonian $\widehat{T}_{1}+\widehat{T}_{2}=\widehat{H}_{\text{N}}+\frac{1}{3}(k_1^2+2k_2^2)$ 
where $\widehat{H}_{\text{N}}=p_1^2+p_2^2+p_3^2+V_{\text{N}}$ with 
\begin{align}
\widehat{T}_{1}&=L_1^2+2L_3^2-\Ti \{L_2 ,L_3\}+V_{\text{N}}{+}\frac{4k_2^2}{(s_2+\Ti s_3)^2}-\frac{4k_2(s_1k_3+2s_2k_2)}{(s_2+\Ti s_3)}+\frac{k_1^2+4k_1k_2+14k_2^2}{3} \, , \\
\widehat{T}_{2}&= (L_2+\Ti L_3)^2-4 s_1k_2\frac{k_2 s_1-k_3 (s_2+\Ti s_3)}{(s_2+\Ti s_3){}^2}
 -\frac{4}{3} k_1 k_2 \,, \\
 \widehat{T}_{3}&=\frac{1}{2}\{L_2+\Ti L_3,L_1\}-s_1(s_2+\Ti s_3)V_{\text{N}}+\frac{s_1 k_2^2}{(s_2+\Ti s_3)^3}-\frac{k_2 k_3}{(s_2+\Ti s_3)^2}+\frac{k_3(k_1+3k_2)}{3}\, .
\end{align}
Both families of Hamiltonians $\widehat{H}_{a,b}$ and $\widehat{H}_{\text{N}}$ posses real spectra by virtue of the $\mathcal{PT}$-symmetry  (\ref{PTtra}). In particular, $\widehat{H}_{0,\frac{1}{2}}$,  $\widehat{H}_{1, \frac{1}{2}}$ and $\widehat{H}_{\text{N}}$ can be also obtained from the systems in the hyperboloid with Minkowski metric signature under the reduction scheme \cite{InSys2} followed with the complex transformation $s_3\rightarrow \Ti s_3$ and
$p_3\rightarrow -\Ti p_3$. The examples described here are just a sample of a larger class of complex models that can be constructed by more generic MASAs
\begin{align}\label{zgen}
Z_\mu=c_{\mu}^0 X_0+c_\mu^1 X_1+c_\mu^2 X_2+c_\mu^4 X_4+c_\mu^6 X_6+c_\mu^8 X_8 \, ,\quad \mu=1,2,3.
\end{align}
after the necessary constraints on the coefficients $c_\mu^\nu$ related with $[Z_\mu,Z_\nu]=0$. The freedom in the number of coefficients, complex or real, allows to construct potentials such as (\ref{potone}), (\ref{potm}) and (\ref{spe1}) that connect continuously systems from different MASAs and separable coordinates \cite{InSys2, comp2}.

\section{Discussion}
\label{SecCon}

In this article, the Marsden--Weinstein reduction method has been extended for the construction of complex $\mathcal{PT}$-symmetric superintegrable Hamiltonians in arbitrary dimensions. Relaxing the reality condition in the linear combinations of the MASAs expands the number of free parameters and thus the variety of the potentials. Since this idea works for arbitrary dimensions and the number of parameters in algebras like $su(n)$ grows with it, many new systems are waiting to be found. In the non-Hermitian scenario, it is possible to design superintegrable systems with real energies with additional parameters, on top of the $n$ couplings coming from the phase space reduction vector $p_x=k$. These extra parameters reveal features such as spontaneous $\mathcal{PT}$-symmetry breaking since the wave functions of the separable coordinates depend explicitly on them. To get a taste of how the complex mechanism works, different classes of systems in one and two dimensions have been discussed, covering complex extensions to known Hermitian Hamiltonians and generalisations of complex Hamiltonians classified by their separable properties  \cite{InSys1,InSys2, comp2, comp3, comp4}. The search and design of non-Hermitian superintegrable systems for three and higher dimensions with this method remains as an open problem. The next natural question is how to appropriately classify such systems, which can be approached in two different but related ways. The difference between the real and the complex case is that, on the algebraic level, the real MASAs are equivalent to each other if they are related by the adjoint action $g Y g^{-1}$ of the corresponding group \cite{InSys1}. With additional complex parameters beyond the standard coupling constants, it is necessary to study how these equivalence classes are defined in the non-Hermitian case. Another way to understand the nature of non-Hermitian Hamiltonians follows from the classification and superintegrability in terms of separable coordinates and algebraic structures  \cite{recent}. In this sense it is interesting to see how systems like the two-parametric potential (\ref{spe1}), fit into this scheme. So far, the solutions of the energy  levels have been  discussed in terms of the Schr\"odinger equation, where the Hamiltonian is the Casimir operator of the corresponding integrals. One may wonder how other tools for obtaining the spectrum can be applied in this non-Hermitian scenario. A very successful technique that works in the real case and can be applied to the complex one is based on parabosonic algebras \cite{Dask,mar14}. This algebraic method was applied in particular to Hamiltonians involving Painlev\'e transcendents \cite{mar09}, where other usual approaches to partial and ordinary differential equations (such as series solutions) were not applicable to obtain the spectrum and the wave functions of the corresponding Schr\"odinger equations. These cases suggested the need to develop further algebraic methods in the context of quantum Hamiltonians. However, this article does not rely on parabosonic algebras, the algebraic scheme is purely based on an underlying $u(3)$ Lie algebra and its enveloping algebra. Such a connection with a Lie algebra then provides properties such as analytic solvability and superintegrability of various models described here. In fact, some of them lead to special functions/orthogonal polynomials beyond the scope of the usual hypergeometric functions or Jacobi polynomials \cite{JacobiOr}. It is known that the Hermitian generic superintegrable systems on the $n$-sphere are related to Lauricella polynomials, which in turn arise from differential equations generalising hypergeometric ones  \cite{miller15,iliev18}. This shows how the programme started in this article on $\mathcal{PT}$-symmetric (and more generally non-Hermitian) superintegrable systems may lead to further generalisations of these special functions, which can potentially be applied beyond quantum systems. Another interesting algebraic aspect was revealed in the paper \cite{miller15}, where a hidden symmetry extended the generic model on the $n$-sphere to quasi exact solvability. This may also lead to an algebra of hidden symmetry for the models proposed here and possible ways to study new complex extensions of quasi exact solvable systems. Finally, the present work provides further evidence for the conjecture that superintegrable models can be obtained from reductions of free particles in higher-dimensional spaces, it also holds for non-Hermitian Hamiltonians. In this sense, the existence of non-Hermitian integrable systems with higher-order conserved quantities and their relation to other reductions is also intriguing, see \cite{thenew} and references therein.

\section*{Acknowledgement}
FC and LI were supported by Fondecyt grants 1211356 and 3220327, respectively. IM was supported by  Australian Research Council Future Fellowship FT180100099. 
\appendix
\section{Explicit computation of the reduced Hamiltonian}
\label{AppA}
This appendix gives some details on the calculation of the complex reduced Hamiltonian in the sphere, where the real case appears as a particular one and can be found explicitly in \cite{InSys1}. The construction of the Hamiltonian (\ref{genrealcase}) requires the Jacobian of the transformation (\ref{transa}) and its inverse, which are given by 
\begin{equation}
J=\frac{\partial(q,r)}{\partial (x,s)}=\left(\begin{array}{cc}
A &B \\
-(B^{-1})^2A & B^{-1}
\end{array}\right)\,,\qquad
J^{-1}=\frac{1}{2}\left(\begin{array}{cc}
A^{-1} &-A^{-1}B^2 \\
B^{-1} & B
\end{array}\right)\,,
\end{equation}
where the components of the matrix $A$ are defined in the following form, 
 \begin{equation}
\label{A}
A_{\mu\nu}(x)=\frac{\partial q_\mu}{\partial x_\nu}=({Z}_\nu)_{\mu\sigma}B_{\sigma\rho}(x)s_\rho=
B_{\mu\rho}(x)({Z}_\nu)_{\rho\sigma}s_\sigma\,. 
\end{equation}
The inverse Jacobian satisfies the equation, 
\begin{eqnarray} \label{jeq}
\label{JIJ}
J^{-1}\mathcal{I}(J^{-1})^T=\frac{1}{2}\left(
\begin{array}{ll}
\mathcal{V}^{-1} & 0\\
0 & \I
\end{array}
\right)\,,\qquad
\mathcal{I}=\left(\begin{array}{cc}
0 & \I\\
\I & 0
\end{array}\right)\,,
\end{eqnarray}
where the matrix $\mathcal{V}$, the same as in Eq. (\ref{ComV}), is independent of the $x$ coordinates
\begin{equation}
\mathcal{V}=-A^T (B^{-1})^2 A=-(B^{-1} A)^T(B^{-1}A)=-(A(x=0))^T(A(x=0)\,.
\end{equation}
The final form of the Hamiltonian is then directly computed with (\ref{jeq}), 
\begin{align}
\begin{array}{lll}
H&=4p_q^Tp_r=(p_q^T,p_r^T)
(2\mathcal{I})\left(\begin{array}{c}
p_q\\
p_r
\end{array}\right)=
(p_x^T,p_s^T)J^{-1}(2\mathcal{I})
(J^{-1})^T\left(\begin{array}{c}
p_x\\
p_s
\end{array}\right)\\
&= p_s^Tp_s+p_x^T \mathcal{V}^{-1} p_x\,.
\end{array}
\end{align}

\end{document}